\input{aipcheck}

\documentclass[final]{aipproc}

\usepackage{color}
\usepackage{amsmath}

\layoutstyle{6x9}

\begin{document}

\title{An Analysis Of $\gamma p \rightarrow p \pi^+ \pi^-$ Using The CLAS Detector}

\author{Matthew Bellis}{
  address={Rensselaer Polytechnic Institute, Troy, NY 12180}
}

\author{CLAS Collaboration}{
  address={Jefferson Lab, Newport News, VA}
}

\begin{abstract}
 Two charged pion final states are studied for photons incident
 on protons. The data come from Thomas Jefferson National Accelerator Facility using the CLAS detector.
 A tagged photon beam of 0.5-2.4 GeV/c was produced through bremsstrahlung radiation and
 was incident on a $\ell$H$_2$ target. This analysis looks at the reaction
 $\gamma p \rightarrow p \pi^+ \pi^-$ using a partial wave analyis to identify
 intermediate baryon resonances. Total cross section is compared to previous experiments
 and preliminary differential cross sections for intermediate baryon resonance quantum numbers are shown.
\end{abstract}

\maketitle

%%%%%%%%%%%%%%%%%%%%%%%%%%%%%%%%%%%%%%%%%%%%
%% MAINMATTER
%%%%%%%%%%%%%%%%%%%%%%%%%%%%%%%%%%%%%%%%%%%%

\section{Introduction}

The consituent quark model does an excellent job of predicting the spectrum
of the majority of baryons and mesons. Capstick, Isgur and others\cite{Koniuk:1980vy,Capstick:1993th,Capstick:1994kb,Capstick:1986bm}
have augmented the quark model for baryons, including decays,
with QCD-inspired corrections and get very good
agreement with experimental data.

But it has been known since the 1960's that there are predicted baryon resonances
which are not observed in the experimental data\cite{Faiman:1968js,Faiman:1969at}. Many of the models use a harmonic 
oscilliator basis, and it is found that these missing states all fall in the
N=2 band. This prompted Lichtenberg\cite{Lichtenberg:1969pp} to propose the diquark model, where 
two of the three quarks become tightly bound, reducing the number of degrees of 
freedom. This constraint leads to a spectrum devoid of the missing resonances
of the full model. There is nothing in QCD however, which would imply any sort of
diquark coupling. Later calculations \cite{Koniuk:1980vy,Forsyth:1983dq,Capstick:1993th} suggest that
these missing states may couple more strongly to $N\pi\pi$ final states than
$N\pi$ final states. As the majority of relevant experiments involve $N\pi$ scattering,
it may not be that surprising that we have not observed these missing resonances. 
JLab is in an excellent position to supplement the world's data with a large
data set of $N\gamma$ scattering. 

We perform a partial wave analysis on the reaction $\gamma p \rightarrow p \pi^+ \pi^-$.
By extracting the partial wave amplitudes it is hoped that any missing baryon states can be identified.
Both intensity and relative phases should give us the handle needed to indentify resonant
states. In addition, this technique gives us the best description of the data and allows us
to accurately calculate both the total and differential cross section. We show that
this technique works and even in our preliminary analysis there appear to be some promising
signals in the data.

\section{The CLAS Detector and Data Selection}

The data was collected at the CLAS (CEBAF Large Acceptance Spectrometer)
at Jefferson Lab in Newport News, VA. This sample of data is taken from
the ``g1c" running period which ran from Oct.-Nov. 1999. About 15\% of the total
run period is analyzed. 

A 2.445 GeV electron beam was directed onto a thin foil radiator to 
produce a bremsstrahlung photon beam. CLAS is equipped with a hodoscope which allows
tagging of photons with energies between 20\% and 95\% of the electron beam.
This corresponds to a center-of-mass $W$ from 1.3 to 2.3 GeV/c$^2$.

CLAS contains a large toroidal magnet to for determining the momentum
of charged particles. A time-of-flight system is used for particle identification\cite{Mecking:2003zu}.
The particle identification is very clean for pions and protons. By making cuts on missing mass
and missing $z$-component of momentum, we are able to identify exclusive events. 

To simulate the detector a {\sc geant}-based program was used. A detailed study of the acceptance
was performed to check the agreement between the simulation and real-world data. 
This was used to identify our fiducial cuts. In the end, we have a very clean sample of
775,553 exclusive events.

\section{Partial Wave Analysis}

\begin{figure}
	\label{decomp}
  \includegraphics[height=.08\textheight]{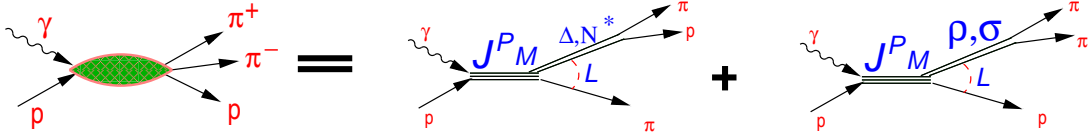}
  \caption{Representation of the decomposition of the scattering
	process into individual partial waves.}
\end{figure}

The purpose of this analysis is to extract the partial waves amplitudes for this
reaction. We want to expand the amplitude in some basis. For this study
we use an $s$-channel decay basis, where we assume that all decays proceed through
2-body decays\cite{Jacob:1959at,Chung:1971ri}. 
Fig.~\ref{decomp} is a representation 
of how we label our basis states: $J,P$ and $M$ of the intermediate state, and
the quantum numbers of the subsequent decays. We also sum over the initial and
final state helicities and add them incoherently. 

Slightly more formally we can represent the $T$-matrix in the following 
fashion.
\begin{eqnarray*}
T_{fi} &=& \langle p\pi^+\pi^-; \tau_f|T|\gamma p; E\rangle\\
       &=& \textcolor{red}{\sum_{\alpha}}\langle p\pi^+\pi^-; \tau_f\textcolor{red}{|\alpha\rangle\langle\alpha|}T|\gamma p; E\rangle\\
       &=& \textcolor{red}{\sum_{\alpha}}\psi^{\textcolor{red}{\alpha}}(\tau_f)V^{\textcolor{red}{\alpha}}(E)
\end{eqnarray*}
We can calculate the decay amplitudes, $\psi^{\textcolor{red}{\alpha}}(\tau_f)$ and allow the fit to determine the
production amplitudes, $V^{\textcolor{red}{\alpha}}(E)$. We use $\alpha$ to represent the quantum numbers of the intermediate
``waves" and $\tau$ represents the kinematics of the reaction. 
This is an example of how our decay amplitudes are labeled.

		 $$J^P,M=\frac{1}{2}^+,+\frac{1}{2}\to\left[\Delta^{++}\pi^-\right]_{\ell=1},\lambda_{p_f}=+\frac{1}{2}$$

We remove the energy dependance of the production amplitudes by binning in $W$, the mass of the intermediate state. 
This method allows us to do an {\em energy independent} study of the scattering amplitudes. 

As a starting point for fitting the production amplitudes we need some set of waves. The number of waves
can quickly grow and become too cumbersome for the fitting routines. We started with waves
that had been seen in a previous analysis of $\pi N \rightarrow N \pi \pi$\cite{Manley:1992yb}. These are
shown in Table~\ref{tab:waves}. This list represents 35 production amplitudes that we fit.
We also include a non-interfering amplitude which represents $t$-channel $\rho$ production.

\begin{table}
\begin{tabular}{ccl}
	\hline
	\tablehead{1}{c}{b}{$J^P$} & \tablehead{1}{c}{b}{$M$} & \tablehead{1}{c}{b}{Isobars}\\
	\hline
	$1/2^+$ & $1/2$        & $\Delta\pi$ $\left(\equiv\{\Delta^{++}\pi^-,\Delta^0\pi^+\}\right)$\\
	\hline
	$1/2^-$ & $1/2$        & $\Delta\pi$, $\left(p\rho\right)_{(s=1/2)}$\\
	\hline
	$3/2^+$ & $1/2$, $3/2$ & $\left(\Delta\pi\right)_{(\ell=1)}$, $\left(p\rho\right)_{(s=1/2)}$,
	$\left(p\rho\right)_{(s=3/2; \ell=1,3)}$, $N^\star(1440)\pi$\\
	\hline
	$3/2^-$ & $1/2$, $3/2$ & $\left(\Delta\pi\right)_{(\ell=0,2)}$\\
	\hline
	$5/2^+$ & $1/2$, $3/2$ & $\left(\Delta\pi\right)_{(\ell=1)}$, $p\sigma$\\
	\hline
	$5/2^-$ & $1/2$, $3/2$ & $\left(\Delta\pi\right)_{(\ell=2)}$\\
	\hline
\end{tabular}
\caption{List of $s$-channel waves used in fit.}
\label{tab:waves}
\end{table}

\begin{figure}
	\includegraphics[height=.25\textheight]{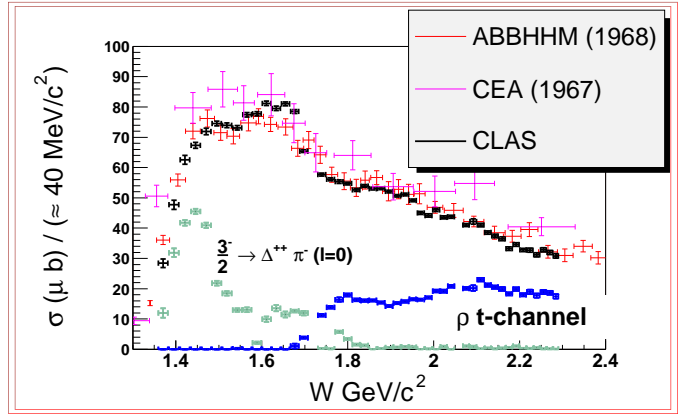}
  \caption{The total cross section for $\gamma p \rightarrow p \pi^+ \pi^-$
	is plotted for this analysis, the CEA\cite{CEA1:1967} experiment, and 
	the ABBHHM\cite{ABBHHM1:1968ke} experiment.}
	\label{cs}
\end{figure}

This procedure allows us to perform a very good acceptance correction and so 
calculate a total cross section. Fig.~\ref{cs} shows our calculated 
cross section plotted vs the results from the ABBHHM collaboration\cite{ABBHHM1:1968ke} as
well as the CEA collaboration\cite{CEA1:1967}. Their results are consistent with ours.
In addition we plot the two strongest waves in the low and high mass region. In the low
mass region, the reaction is dominated by $\frac{3}{2}^-\rightarrow \Delta^{++}\pi^-(\ell = 0)$.
This is to be expected, as some expect the ``contact term" to show up in this wave\cite{Murphy:1996ms}. The high
mass region is dominated by our $t$-channel $\rho$ wave. 

\begin{figure}
	\includegraphics[width=6cm]{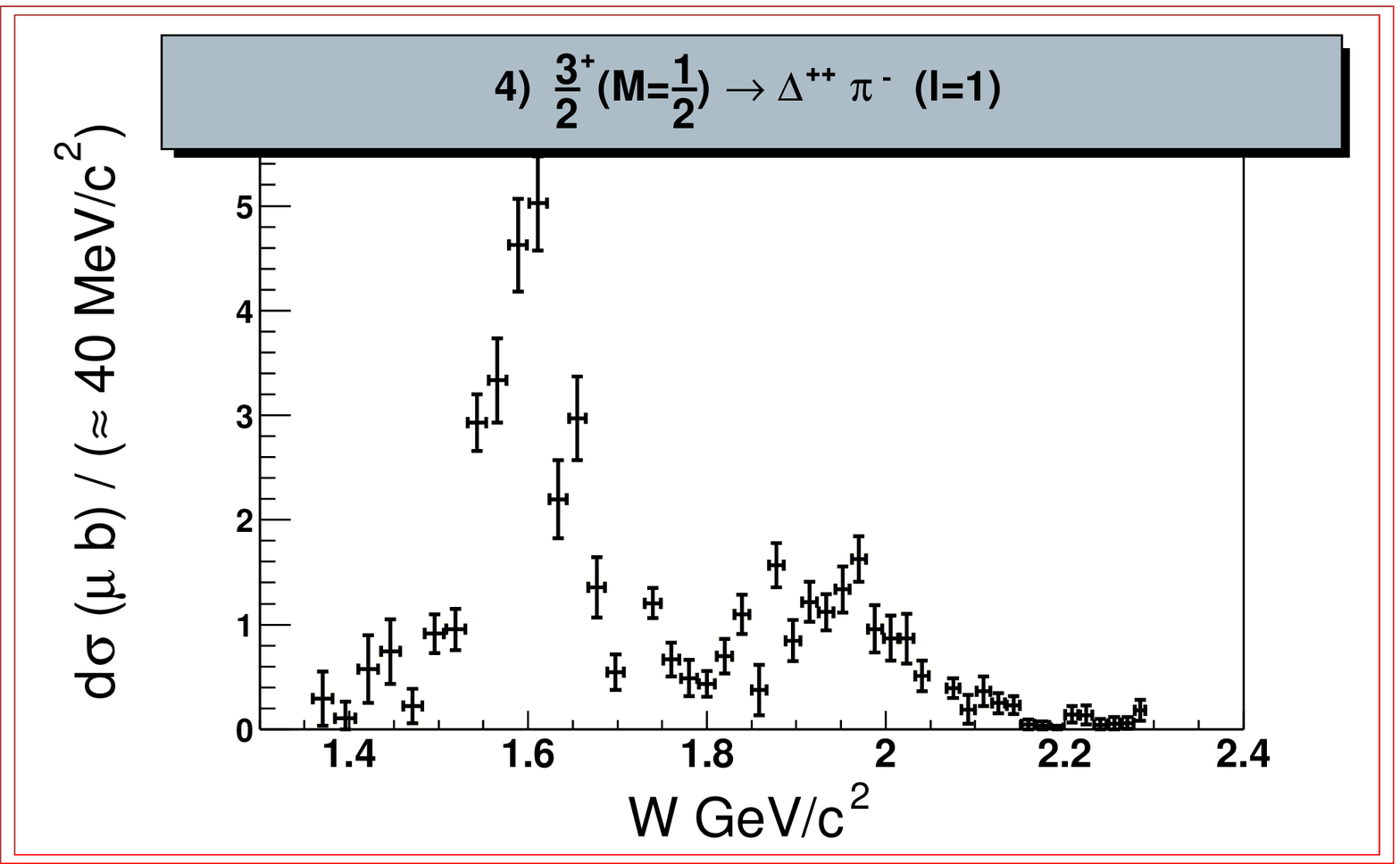}
	\includegraphics[width=6cm]{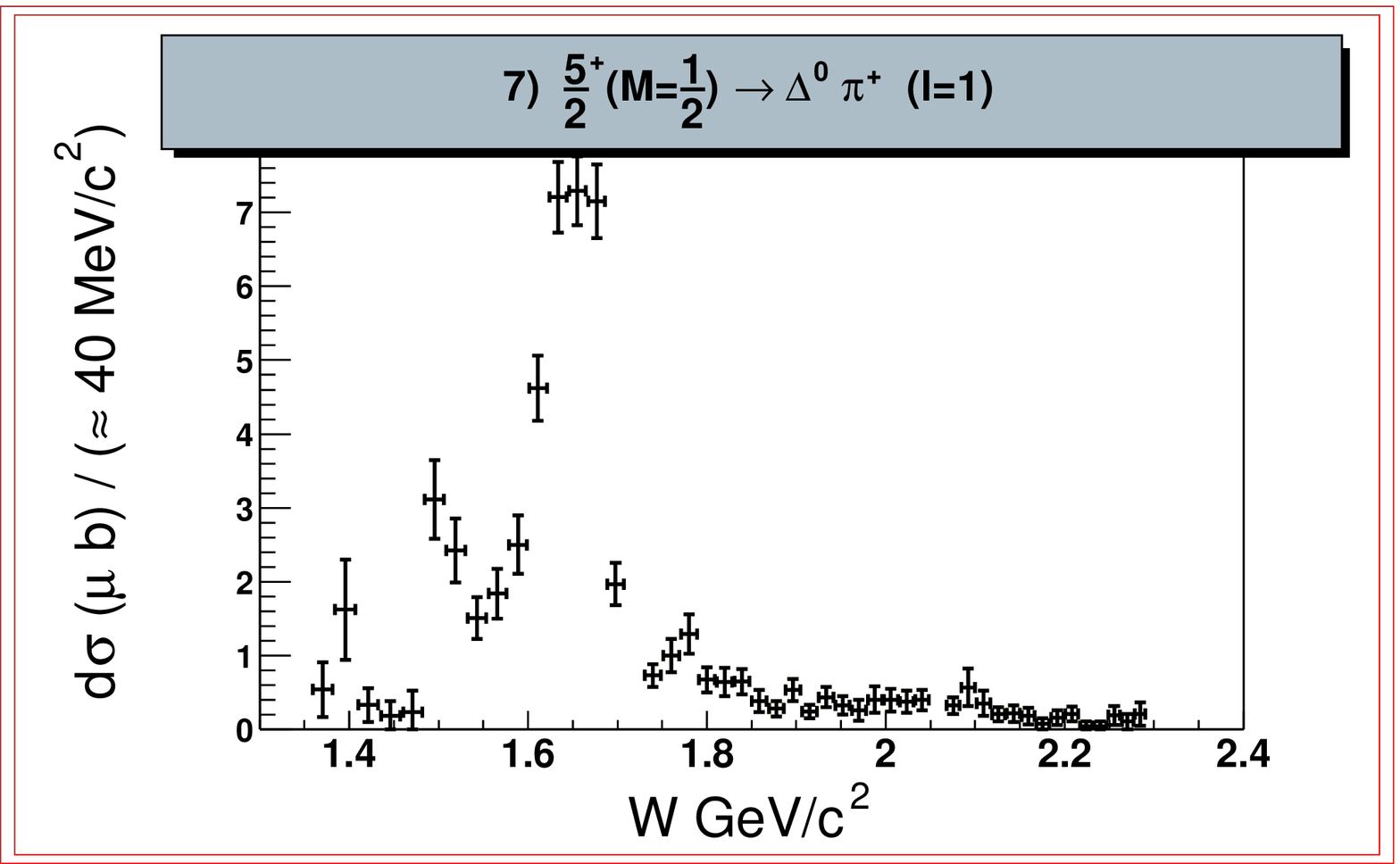}
	\includegraphics[width=6cm]{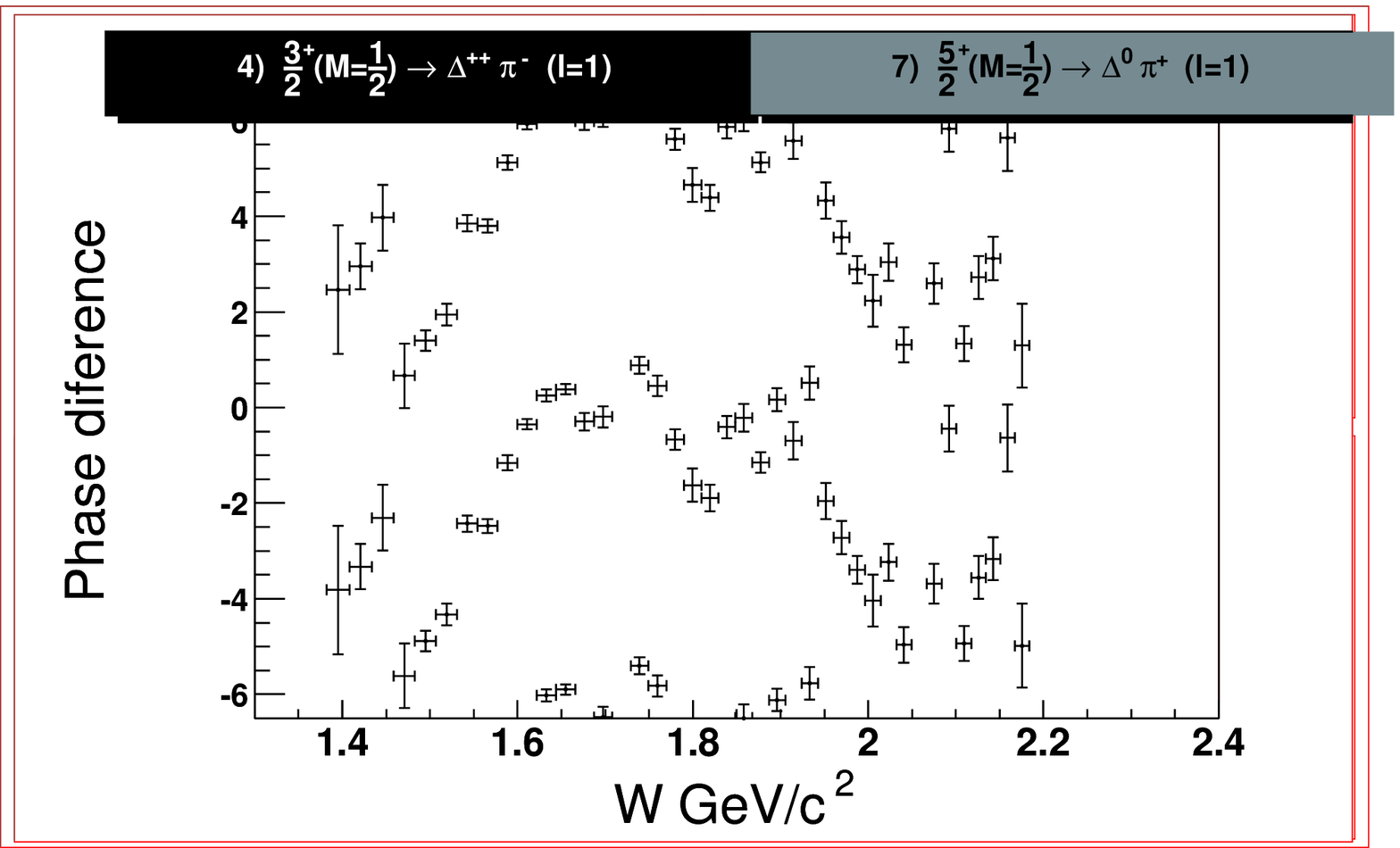}
  \caption{The intensities for two individual partial waves
	are plotted along with 
	the relative phase for the same two partial waves. Beacause their
	phase difference is just an angle, it is plotted 3 times: $\phi$, 
	$\phi+2\pi$ and $\pi-2\pi$.} 
	\label{intpd}
\end{figure}

We can look at individual wave intensities to see if there is evidence of resonance
behavior. Fig.~\ref{intpd} shows the intensities of two waves and their relative phase. 
The $\frac{3}{2}^+$ wave shows an enhancement around 1600 MeV, while the $\frac{5}{2}^+$
wave shows an enhancement around 1650 MeV. The phase motion appears to qualitatively support
the idea that these are resonant waves. One must remember that {\em each point is the result
of an independant fit}. 
We are encouraged that the data seems to be consistent with 
some known states: $\Delta$(1600)$P_{33}$ and N(1680)$F_{15}$. 
In the future, we will explore different wave sets to find the best, stable
description of the physics and perform a full mass dependant analysis. 

The authors would like to thank the organizers of this conference for the opportunity
to present our work. 

\bibliographystyle{aipproc}   % if natbib is available

\bibliography{bib}

\IfFileExists{\jobname.bbl}{}
%\IfFileExists{bib.bbl}{}
 {\typeout{}
  \typeout{******************************************}
  \typeout{** Please run "bibtex \jobname" to optain}
  \typeout{** the bibliography and then re-run LaTeX}
  \typeout{** twice to fix the references!}
  \typeout{******************************************}
  \typeout{}
 }

\end{document}